\newcommand{\dangle}{$\cos\theta_\mathrm{CM}^{d}$}
\documentclass[final,5p,times]{elsarticle}

\usepackage{amssymb}
\usepackage{graphics}
\usepackage{graphicx}
\usepackage{xcolor}
\usepackage{sidecap}
\usepackage{microtype}
\usepackage{multirow}

\usepackage{lineno}

\journal{Physics Letters B}

\begin{document}

\begin{frontmatter}



\title{Coherent  $\pi^0\eta d$ photoproduction at forward deuteron angles measured at BGOOD}
\author[1]{A.J.~Clara Figueiredo}
\ead{figueiredo@physik.uni-bonn.de}
\author[1]{T.C.~Jude}
\ead{jude@physik.uni-bonn.de}
\author[1]{S.~Alef\fnref{fn1}}
\author[2]{P.L.~Cole}
\author[3]{R.~Di Salvo}
\author[1]{D.~Elsner}
\author[3,4]{A.~Fantini}
\author[1]{O.~Freyermuth}
\author[1]{F.~Frommberger}
\author[5,6]{F.~Ghio}
\author[1]{J. Gro\ss}
\author[1]{K.~Kohl}
\author[7]{P.~Levi Sandri}
\author[8,9]{G.~Mandaglio}
\author[10]{P.~Pedroni}
\author[1]{B.-E.~Reitz \fnref{fn1} }%
\author[3,11]{M.~Romaniuk}
\author[1]{G.~Scheluchin\fnref{fn1}}
\author[1]{H.~Schmieden}
\author[1]{A. Sonnenschein}
\author[1]{C. Tillmanns}

\fntext[fn1]{No longer employed in academia}

\address[1]{Rheinische Friedrich-Wilhelms-Universit\"at Bonn, Physikalisches Institut, Nu\ss allee 12, 53115 Bonn, Germany}
\address[2]{Lamar University, Department of Physics, Beaumont, Texas, 77710, USA}
\address[3]{INFN Roma ``Tor Vergata", Via della Ricerca Scientifica 1, 00133, Rome, Italy}
\address[4]{Universit\`a di Roma ``Tor Vergata'', Dipartimento di Fisica, Via della Ricerca Scientifica 1, 00133, Rome, Italy}
\address[5]{INFN sezione di Roma La Sapienza, P.le Aldo Moro 2, 00185, Rome, Italy}
\address[6]{Istituto Superiore di Sanit\`a, Viale Regina Elena 299, 00161, Rome, Italy}

\address[7]{INFN - Laboratori Nazionali di Frascati, Via E. Fermi 54, 00044, Frascati, Italy}
\address[8]{INFN sezione Catania, 95129, Catania, Italy}
\address[9]{Universit\`a degli Studi di Messina, Dipartimento MIFT,  Via F. S. D'Alcontres 31, 98166, Messina, Italy}
\address[10]{INFN sezione di Pavia, Via Agostino Bassi, 6 - 27100 Pavia, Italy}
\address[11]{Institute for Nuclear Research of NASU, 03028, Kyiv, Ukraine}

\begin{abstract}
	
		The coherent reaction, $\gamma d \rightarrow \pi^0\eta d$ was studied with the BGOOD experiment at ELSA from threshold to a centre-of-mass energy of 3200\,MeV.  A full kinematic reconstruction was made, with final state deuterons identified in the forward spectrometer and $\pi^0$ and $\eta$ decays in the central BGO Rugby Ball.  The strength of the differential cross section  exceeds what can be  described by models of coherent photoproduction at forward angles by orders of magnitude.  The distribution of the differential cross section has an excellent agreement with a model including quasi-free $\Delta \pi$ photoproduction, pion re-scattering and $N(1535)$ formation and subsequent nucleon coalescence to the deuteron.  This also gives a reasonable description of the two-body invariant mass distributions and naturally explains the similar magnitudes of this channel and $\pi^0\pi^0 d$ coherent photoproduction.   
\end{abstract}




\end{frontmatter}


\section{Introduction}

$\eta\pi^0$ photoproduction  is a unique reaction to probe hadronic interactions.  Off the nucleon, the channel can access intermediate $N^*$ and $\Delta$ resonances not observed in single meson photoproduction~\cite{klempt10,horn08}, and given the higher centre-of-mass energy can also probe the third and fourth resonance regions.  There is also strong evidence of structure in the $p\eta$ invariant mass originating from a triangle singularity driven by $a_0(980)$~\cite{metag21}.  
The reaction has now been extensively studied both experimentally~\cite{horn08,ajaka08,horn08b,kashevarov09,gutz10,gutz14,sokhoyan18} and with phenomenological models~\cite{fix10,egorov13,egorov20,doering10,debastiani17}, and it is understood to be dominated by the sequential decay, $\gamma N \rightarrow \Delta(1700)\frac{3}{2}^-\rightarrow N(1535)\frac{1}{2}^-\pi^0\rightarrow \pi^0\eta N$, where the $\Delta(1700)$ is dynamically generated from interactions of pseudoscalar mesons and the ground state baryon decuplet~\cite{doering06,kolomeitsev04,sarkar05}.  A detailed knowledge of the reaction off the nucleon suggests that coherent photoproduction off the deuteron should also be well understood, as the elementary amplitudes from the proton and neutron are approximately equal and can be summed coherently due to the isoscalar nature of the deuteron.  This has been the basis for models of coherent photoproduction,  assuming the impulse approximation and including the deuteron momentum form factor (see, for example Ref.~\cite{egorov13}).  Such models give good agreement to  total cross section measurements in Refs.~\cite{kaeser16,ishikawa21,ishikawa22}.

Coherent $\pi^0 \eta$ photoproduction off the deuteron may also be particularly sensitive to $\eta N$ interactions.  The three-body final state permits kinematics with low relative momentum between the $\eta$ and the deuteron which cannot be achieved with single $\eta$ photoproduction.  Parameters, such as the scattering length, $a_{\eta N}$ and mass and widths of a bound system can in principle be determined from invariant mass distributions to characterise $\eta N$ low energy dynamics.  This was a significant motivation for studies with the FOREST detector at the ELPH facility~\cite{ishikawa21,ishikawa22}, where it was suggested that the data supported two sequential mechanisms:  
\begin{eqnarray}
\gamma d \rightarrow  \mathcal{D}_\mathrm{IV}\rightarrow \pi^0 \mathcal{D}_\mathrm{IS} \rightarrow \pi^0\eta d \label{eq:reaction1}\\
\gamma d \rightarrow \mathcal{D}_\mathrm{IV}\rightarrow \eta \mathcal{D}_\mathrm{IV}' \rightarrow \pi^0\eta d\label{eq:reaction2}
\end{eqnarray}

\noindent where $\mathcal{D}_\mathrm{IS}$ and $\mathcal{D}_\mathrm{IV}$ are states with baryon number 2 and isoscalar or isovector respectively.  The invariant mass of the $\eta d$ system exhibited a low mass enhancement not described by the model in Ref.~\cite{egorov13}.  Via a phenomenological analysis and combined fits to $\eta d$, $\pi^0 d$ and $\pi^0 \eta$ invariant masses,  a $\eta d$ state was proposed, corresponding to the $\mathcal{D}_\mathrm{IS}$ in Eq.~\ref{eq:reaction1}.  Additionally, a mass of the $\mathcal{D}_\mathrm{IV}$ was determined which was consistent with the proposed $N\Delta$ dibaryon predicted by calculations in Ref.~\cite{gal14} and experimental data in Refs.~\cite{ishikawa17,ishikawa19}.  This was further supported by the determination of a large scattering length, $a_{\eta d}$ for the $\eta d$ system, however it was undetermined whether a virtual or bound $\eta d$ state was responsible.\footnote{This was due to the sign of $a_{\eta\pi}$ not being able to be resolved.  It was  consequently not clear if it was a bound state where the pole of the scattering amplitude resides below threshold on the real axis of the first Riemann sheet of the complex energy plane, or alternatively a virtual state, where the pole is below threshold on the real axis of the second Riemann sheet. See for example, Ref.~\cite{matuschek21} for a description of near-threshold bound and virtual states.}
These interpretations however are in contrast to Ref.~\cite{torres23}.  This calculation was  based upon a  model describing $\gamma N \rightarrow \Delta(1700)\frac{3}{2}^-\rightarrow N(1535)\frac{1}{2}^-\pi^0\rightarrow \pi^0\eta N$, assuming impulse approximations and additional $\pi$ and $\eta$ re-scattering. 
A description of the invariant mass distributions was achievable without including a bound or virtual $\eta d$ state.  
A large discrepancy in the angular distribution of the differential cross section still remains however, where data exhibits an almost flat distribution over all \dangle{}\footnote{\dangle{} is cosine of the centre-of-mass polar angle of the deuteron.} compared to   model calculations which are backward peaked, under estimating the differential cross section at forward \dangle{} by an order of magnitude.  This loss in strength at forward \dangle{} is expected given the strong dependence on the deuteron form factor, where large momentum transfer kinematics are suppressed due to the small internal binding energy.  The inclusion of $\pi^0$ and $\eta$ re-scattering terms into phenomenological models were shown to increase the calculated forward differential cross section by approximately 10\,\%~\cite{torres23}, however not to the extent observed.  This remaining discrepancy between data and model calculations may demonstrate that an as yet unaccounted for mechanism plays a dominant role in the reaction process.

There is an additional motivation for studying $\pi^0\eta d$ which may be related to the unexpected strength of the differential cross section at forward \dangle{}.  
Since the 1960s, spectra of ``dibaryon systems" beyond the deuteron have been suggested, with searches focusing on isovector candidates.  Interpretations of candidate systems ranged from genuine dibaryons to ``box diagrams" of Pion Exchange Models (OPE) between nucleons  (for a recent review see  Ref.~\cite{clement17} and earlier reviews see Refs.~\cite{locher86,strakovsky91}).  There has recently been a resurgence in the study of candidates of bound two baryon systems beyond the deuteron.  This began with the observation of the $d^*(2380)$ dibaryon candidate first observed in the fusion reaction, $p n \rightarrow d\pi^0\pi^0$~\cite{adlarson11,bashkanov09}, and now in a multitude of final states and observables~\cite{adlarson13,adlarson14,adlarson14PRC,bashkanov19,bashkanov20_pgamma,adlarson13PRC,adlarson15PLB}.  If present, coherent photoproduction off the deuteron provides a particularly clean probe of intermediate dibaryon formation as the cross section for  conventional coherent processes is expected to be suppressed due to the large momentum transfer.  The reaction $\gamma d \rightarrow \pi^0 \pi^0 d$ was studied with the FOREST detector at ELPH~\cite{ishikawa17,ishikawa19} where it was argued there was evidence of the $d^*(2380)$ and higher lying isoscalar dibaryons at 2.47 and 2.63\,GeV/c$^2$ and an isovector dibaryon at 2.14\,GeV/c$^2$.  
 The BGOOD collaboration subsequently measured the $\pi^0\pi^0 d$ reaction at forward \dangle~\cite{coherentpaper}, where the differential cross section  was orders of magnitude higher than model calculations~\cite{fix05}.  The BGOOD invariant mass distribution measurements of the two-body systems were consistent with the FOREST detector at ELPH collaboration claims of dibaryon formation, however improved statistical precision would be needed for a conclusive statement.  $\pi^0\eta d$ measurements  would be complementary to the $\pi^0\pi^0 d$ channel and may  shed light on the reaction mechanism at forward \dangle{} and the role of potential dibaryon candidates.  A single intermediate dibaryon state in $\pi^0\pi^0 d$ photoproduction must be isoscalar, whereas isovector systems can also contribute in  $\pi^0\eta d$ photoproduction.  Both channels however would permit sequential decay mechanisms similar to what is proposed in Eqs.~\ref{eq:reaction1} and \ref{eq:reaction2}. 
 
 Understanding the reaction mechanism for $\pi^0\eta d$ coherent photoproduction, particularly at forward \dangle{} is therefore the motivation for the measurements presented here.  The BGOOD photoproduction experiment~\cite{technicalpaper} at the ELSA facility~\cite{hillert06,hillert17} at the University of Bonn is ideally suited to measure coherent reactions, with clean identification of deuterons  produced at forward angles, as demonstrated in Ref.~\cite{coherentpaper} for the $\pi^0\pi^0 d$ channel.  This is essential to distinguish between coherent reactions and quasi-free reactions off the proton with cross sections orders of magnitude higher.

\section{Experimental setup and analysis procedure}

BGOOD is comprised of two main parts: a central calorimeter region, ideal for neutral meson identification, and a magnetic \textit{Forward Spectrometer} for charged particle identification and momentum reconstruction (for a detailed description see Ref.~\cite{technicalpaper}).  The \textit{BGO Rugby Ball} is the main detector over the central region, covering laboratory polar angles 25 to 155$^\circ$.  The detector is comprised of 480 BGO crystals for the reconstruction of photon momenta via electromagnetic showers in the crystals.  The separate time readout per crystal enables a clean separation and identification of neutral meson decays.  Between the BGO Rugby Ball and the target are the \textit{Plastic Scintillating Barrel} for charged particle identification via $\Delta E-E$ techniques and the  \textit {MWPC} for charged particle tracking and vertex reconstruction.

The Forward Spectrometer covers a laboratory polar angle 1-12$^\circ$. The tracking detectors, \textit{MOMO} and \textit{SciFi} are used to track charged particles from the target.  Downstream of these is the \textit{Open Dipole Magnet} operating at an integrated field strength of 0.216\,Tm.  A series of eight double sided \textit{Drift Chambers} track charged particle trajectories after the curvature in the magnetic fields and are used to determine particle momenta with a resolution of approximately 6\,\%.\footnote{The resolution improves to 3\,\% if the Open Dipole Magnet is operating at the maximum integrated field strength of 0.432\,Tm.}  Three \textit{Time of Flight (ToF) Walls} downstream of the drift chambers determine particle $\beta$ and are used in combination with the measured momentum for particle identification via  mass determination.
 Track reconstruction in the Forward Spectrometer is described in Ref.~\cite{technicalpaper}. 

The small intermediate region between the central region and the Forward Spectrometer is covered by \textit{SciRi}, which consists of three concentric rings, each with 32 plastic scintillators for charged particle detection. 

The deuterium target data presented was taken over a period of 26 days using an 11\,cm long target and an ELSA electron beam energy of 2.9\,GeV.
The electron beam was incident upon a thin diamond radiator\footnote{A diamond radiator was used to produce a coherent, linearly polarised photon beam with a maximum polarisation at a beam energy of 1.4\,GeV, however the polarisation was not required for the presented analysis.} to produce an energy tagged bremsstrahlung photon beam which was subsequently collimated.  The photon beam energy, $E_\gamma$, was determined per event by momentum analysing the post bremsstrahlung electrons in the \textit{Photon Tagger}.  The integrated photon flux from $E_\gamma = 800$ to 1600\,MeV (the region of the presented data) was $8.9 \times 10^{12}$.  The hardware trigger required a tagged incident photon and a minimum energy deposition in the BGO Rugby Ball of approximately 150\,MeV (see Ref.~\cite{klambdapaper} for details).  Data using a liquid hydrogen target was used to subtract background from quasi-free reactions off the proton.  This dataset was taken over 19 days with an integrated photon flux from $E_\gamma = 800$ to 1600\,MeV of $4.7 \times 10^{12}$. The hardware and running conditions were identical to the liquid deuterium target data.


Candidate events were selected where exactly four photons were identified in the BGO Rugby Ball (via a veto with the Plastic Scintillating Barrel) and one charged particle in the Forward Spectrometer, corresponding to the two $\pi^0/\eta\rightarrow \gamma\gamma$ decays and a forward going deuteron respectively.  Events were rejected if any additional charged particle was identified.  Events required at least one combination where the invariant mass of two photons were within 40\,MeV/c$^2$ of the $\pi^0$ mass and the other two photons within 80\,MeV/c$^2$ of the $\eta$ mass (corresponding to approximately $2.5\sigma$ of the mass resolution for each meson).  All combinations per event meeting this criterion were retained.  The reconstructed mass of the forward going deuteron was required to be between 1550 to 2500\,MeV/c$^2$, which is 1.3$\sigma$ below and $2.6\sigma$ above the deuteron mass.   This asymmetric selection limited background from quasi-free reactions off the proton.

\begin{figure} [h]
	\centering
	\includegraphics[trim={18cm 1cm 0cm 0cm},clip,width=\columnwidth]{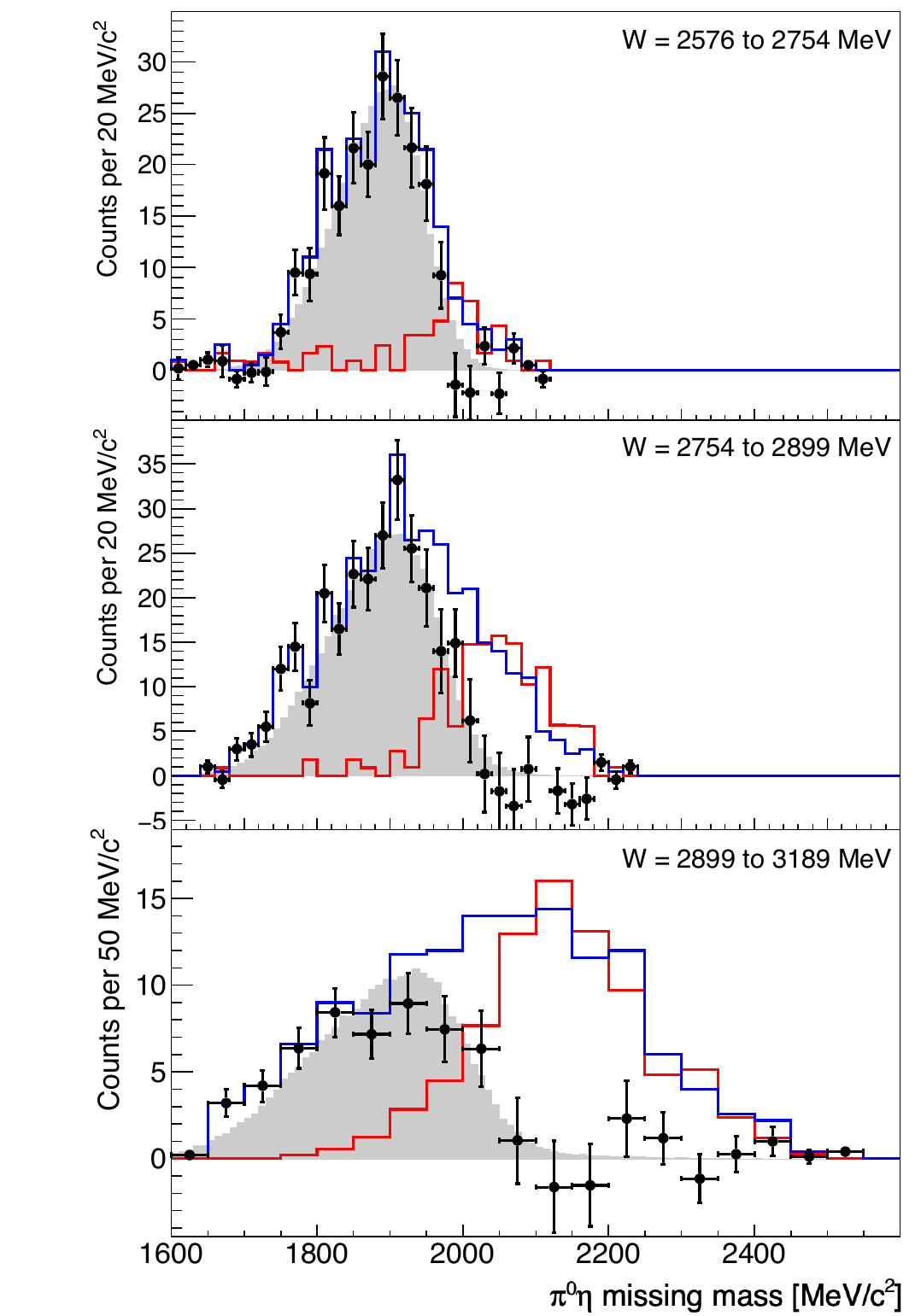}
	\caption{The missing mass recoiling from the $\pi^0\eta$ system for centre-of-mass energy ranges labelled inset.  Deuterium target data is the blue line and hydrogen target data, normalised according to flux is the  red line.  The black circles are the deuterium data with normalised hydrogen data subtracted (vertical error bars indicate statistical error, horizontal error bars are the bin widths).  The grey shaded region is simulated $\pi^0\eta d$, where the integral is equal to the real data after subtraction of the normalised hydrogen data.}		
	\label{fig:missmasstotal}
\end{figure}

Simulated data demonstrated that there was negligible contamination in the event yield from the coherent reactions, $\gamma d \rightarrow \pi^0\pi^0 d$ and $\gamma d \rightarrow \pi^0\pi^0 \pi^0 d$ which have been observed to have comparable differential cross sections at forward deuteron angles~\cite{coherentpaper,andreasthesis}.  Significant background remained however from quasi-free reactions off the proton, where the proton was misidentified as a deuteron in the Forward Spectrometer.
This is  observed in histograms of the missing mass recoiling from the $\pi^0\eta$ system, on the condition that a forward deuteron candidate is also identified (Fig.~\ref{fig:missmasstotal}).
Data using the deuterium target is shown as the blue line.  There is a peak at the expected deuteron mass of 1875.6\,MeV/c$^2$, with a shoulder at higher masses that increases in magnitude with increasing centre-of-mass energy, $W$.
This shoulder originates from misidentified protons in the Forward Spectrometer and resides at higher masses due to the false assignment of the target mass.  	Data using the hydrogen target (normalised to the deuterium dataset photon flux) is shown as the red line and has a good agreement to the magnitude and shape of the high energy shoulder in the deuterium data.  The black data points are the deuterium data with the hydrogen data subtracted, leaving only the coherent reaction contribution.  This has excellent agreement to the simulated $\gamma d \rightarrow \pi^0\eta d$ data shown as the grey shaded region.  The Fermi motion which is only present in the quasi-free reactions off the deuteron target was sufficiently small to make negligible difference to the distributions.

The yield of events for each photon energy interval was subsequently determined from the integral of the subtracted deuterium data.  The integrated range was asymmetric over the deuteron mass peak, from 1600 MeV/c$^{2}$ to $1\sigma$ above the deuteron mass, which increased with photon beam energy.  This asymmetric selection improved the statistical precision by not including bins at higher missing mass with large statistical uncertainty.

Approximately 10\% of events passed selection criteria twice due to the possibility of reconstructing the $\pi^0$ and $\eta$ from two different combinations of the four detected photons.  These events contributed a single count to the differential cross section with respect to $W$.  These events however were rejected for the two-body invariant mass distributions due to this ambiguity.

The detection efficiency was determined using the BGOOD GEANT4~\cite{geant4} simulation, including all spatial, energy and time resolutions, magnetic fields and hardware efficiencies (see Ref.~\cite{technicalpaper} for details).  
Unlike a two-body final state, the kinematics of three final state particles are not fixed by $W$ and \dangle{}, and the distributions of which may change the measured detection efficiency.  This was investigated by iteratively changing two different event distributions, either initially a phase space distribution, or assuming a sequential reaction mechanism $\gamma d \rightarrow NN(1535) \pi^0 \rightarrow \pi^0\eta d$ shown in Fig.~\ref{fig:coherentdiagrams}(d).  This assumes a bound $NN(1535)$, where the mass of the $N(1535)$ was event-wise sampled from the expected Breit-Wigner width and the bound system subsequently decays to $d\eta$.  The emphasise was not to test a specific reaction mechanism but rather to ensure two  different distributions were used as starting points to determine systematic uncertainties.
Differential cross sections with respect to the invariant mass of the two-body systems ($\eta\pi^0$, $\pi^0 d$ and $\eta d$) were determined using detection efficiencies from  both distributions.  The measured differential cross section distributions were then iteratively included as the next distributions for the event generators used to determine the detection efficiency.  This was performed six times, until negligible changes were apparent either from the previous iteration or between the final iterations of the two different starting distributions.


  Systematic uncertainties from the forward track finding  (1.0\,\%), timing cuts (2.0\,\%), beam spot alignment  (4.0\,\%) and different sub-detector efficiencies were determined previously~\cite{klambdapaper}.  From varying selection cuts by $\pm 10$\,\%, the systematic uncertainties of identifying $\pi^0$, $\eta$ and the deuteron were estimated as 1.3, 2.5 and 1.0\,\% respectively.  The uncertainty in the detection efficiency was determined as 2\,\% by comparing the last two  iterations of the 3-body final state distributions.  When summed in quadrature,  this gave a total systematic uncertainty of 9\,\%.
  
  \section{Results and interpretation}\label{sec:results}
    
  The differential cross section  versus $W$ for \dangle{} $> 0.8$ is shown in Fig.~\ref{fig:cs} as the solid black circles.  The magnitude is similar to the previously measured $\pi^0\pi^0d$ differential  cross section at BGOOD~\cite{coherentpaper} (blue squares), suggesting that in both cases a  reaction mechanism dominates which is not present in reactions off the nucleon, where  $\pi^0\eta N$ is of the order of three times smaller than $\pi^0\pi^0 N$.  The data agree well with the three data points from the previous measurement at the FOREST detector at ELPH~\cite{ishikawa22}, supporting the flat distribution with respect to \dangle{} that was reported.  This is in contrast to model calculations where it is expected that the differential cross section is suppressed at forward \dangle{ due to the large momentum transferred to the deuteron.  For \dangle{} $> 0.8$ and $W = 2600$\,MeV (close to threshold), the three-momentum transfer to the deuteron is approximately 650\,MeV/c which is much higher than the Fermi momentum of the constituent nucleons (typically 80\,MeV/c) and therefore what can be transferred to the deuteron for it to remain intact.  The three momentum transfer increases approximately linearly, where at $W = 3$\,GeV it is approximately 1460\,MeV/c.  This is even higher than for  $\pi^0\pi^0 d$ photoproduction, where the differential cross section was also measured with a  surprisingly high magnitude~\cite{coherentpaper}.  

  \begin{figure}[h]
  	\includegraphics[trim={0cm 0cm 0cm 0cm},clip,width=0.5\textwidth]{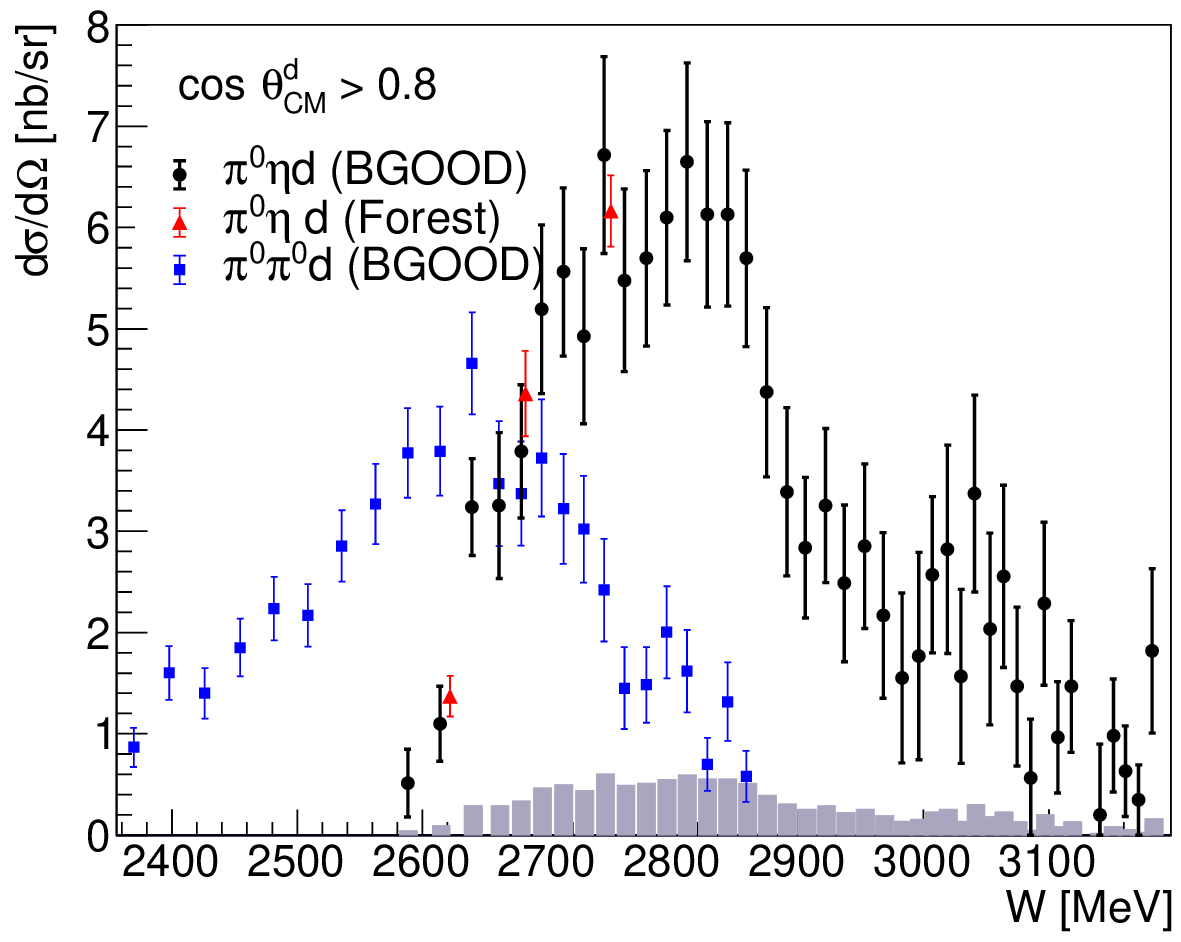}
  	\caption{The $\gamma d \rightarrow \pi^0 \eta d$ differential cross section for \dangle{}$> 0.8$ (filled black circles).  
  		The systematic errors are the grey bars on the abscissa.  Previous data from the FOREST detector at ELPH~\cite{ishikawa22} and shown as the three red triangles.  Previous BGOOD data for $\gamma d \rightarrow \pi^0\pi^0 d$~\cite{coherentpaper} is shown as the filled blue squares for comparison (systematic errors not shown).}
  	\label{fig:cs}
  \end{figure}

Models of pion exchange and re-scattering off nucleons were considered in an attempt to qualitatively understand the distribution of the differential cross section.  Figure~\ref{fig:coherentdiagrams}(a) and (b) show  diagrams of quasi-free reactions, where either a $\Delta$ or $N(1535)$ is produced off the proton in addition to a $\pi^+$ which re-scatters off the neutron to produce either a $N(1535)$ or a $\Delta$.  After their subsequent decays, the nucleons coalesce to a deuteron if there is sufficiently small relative momentum between them.  Such a mechanism could be expected to produce a peak in the cross section around the summed $\Delta$ and $N(1535)$ mass of approximately 2770\,MeV.  
If such mechanisms are responsible, this would naturally explain the similar cross section magnitudes of coherent $\pi^0\pi^0 d$ and $\pi^0 \eta d$ photoproduction as the decay branching ratio of the $N(1535)$ to $N\pi^0$ and $N\eta$ are similar (45 and 50\,\% respectively).  

  \begin{figure}[h]
	\includegraphics[trim={0cm 12cm 0cm 0cm},clip,width=0.5\textwidth]{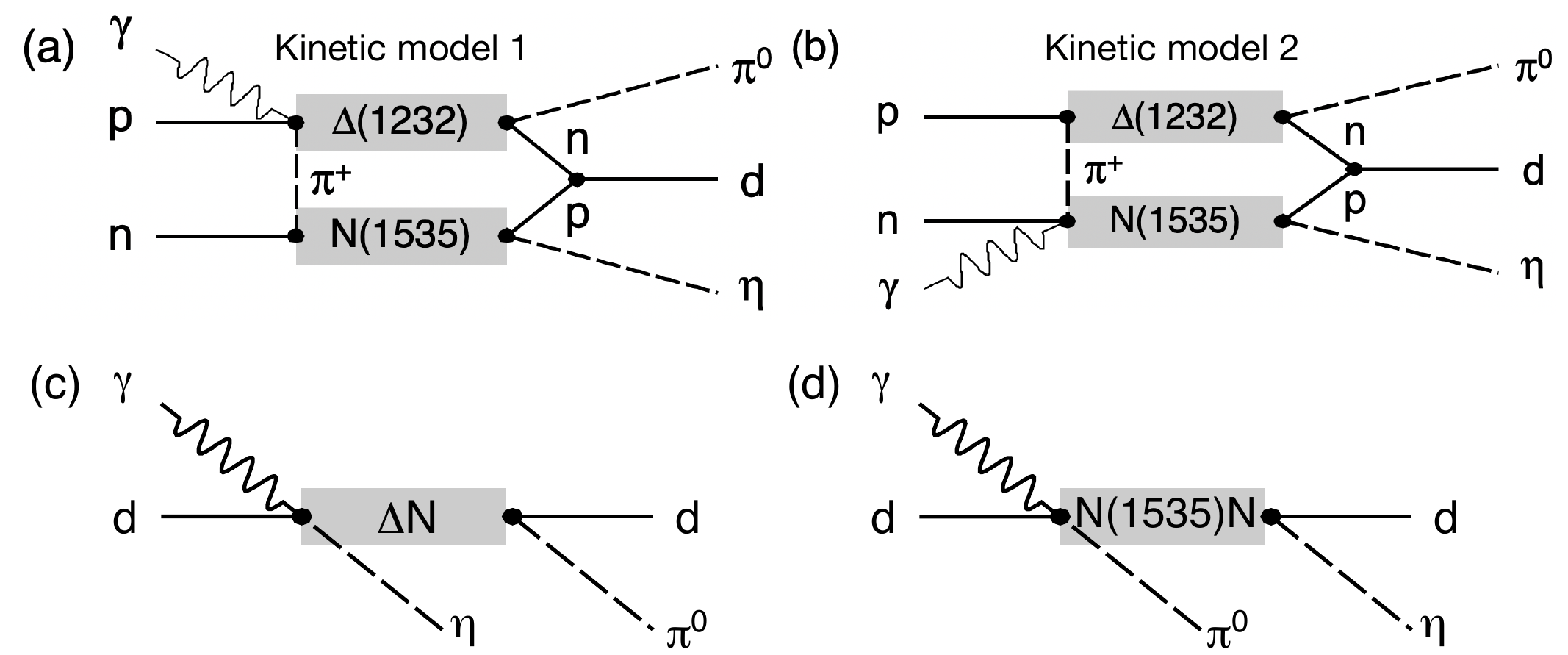}
	\caption{Proposed diagrams contributing to the reaction, $\gamma d \rightarrow \pi^0\eta d$ (described in the text).}
	\label{fig:coherentdiagrams}
\end{figure}
  
Toy models of the two diagrams in Fig.~\ref{fig:coherentdiagrams} were developed, \textit{Kinetic model 1} and  \textit{Kinetic model 2} (denoted KM1 and KM2 herein).  On shell kinematics were assumed at each reaction vertex and phase space decay distributions, with the exception of the $\Delta$ decay, which follows the expected $(1 + 3\cos^2 \theta)$ angular distribution.  The $\Delta$ and $N(1535)$ masses were sampled event-wise from the Breit-Wigner distributions.  The $\pi^+$ re-scattering was weighted by the pion exchange propagator, $1/(m_\pi^2 + q^2)^2$, where $q$ is the $\pi^+$ centre-of-mass momentum in the $\pi^+$ - spectator nucleon system.  An additional $q^2$ weighting was assumed for the dominant magnetic coupling at the $\gamma p\Delta \pi$ vertex.  The final state proton and neutron coalesced to the deuteron if the relative momentum was smaller than the internal Fermi momentum of the deuteron.  This numerical value was randomly chosen event-wise from the expected internal nucleon momentum distribution.
The model is simplistic, however the on-shell kinematics could be expected to give a reasonable approximation near threshold if a single reaction mechanism dominates.

  The amplitudes of the model distributions were determined by a $\chi^2$ minimisation fit to the data, shown in Fig.~\ref{fig:coherentfit}.  It is clear KM1 dominates the differential cross section with an excellent description of the data.  The lower threshold for the $\Delta \pi^+$ photoproduction off the nucleon in KM1 ensures the model agrees with the rise in the spectrum at low $W$ to the peak at approximately 2750\,MeV.  The fit achieved a reduced $\chi^2$ of 1.08, with KM1 contributing over ten times the strength of KM2.  It is interesting  to compare the relative contributions of KM1 and KM2.  The additional $q^2$ dependence of the re-scattered $\pi^+$ at the  $\gamma p\Delta \pi$ vertex
would be expected to significantly reduce the contribution of KM1 compared to KM2 by the order of 180.  However conversely, the expected photoproduction cross section to produce  $\Delta(1232)\pi^+$ for KM1 compared to $N(1535)\pi^+$ for KM2 is of the order of 40 times higher.   



\begin{figure}[h]
	\includegraphics[trim={0cm 0cm 0cm 0cm},clip,width=0.5\textwidth]{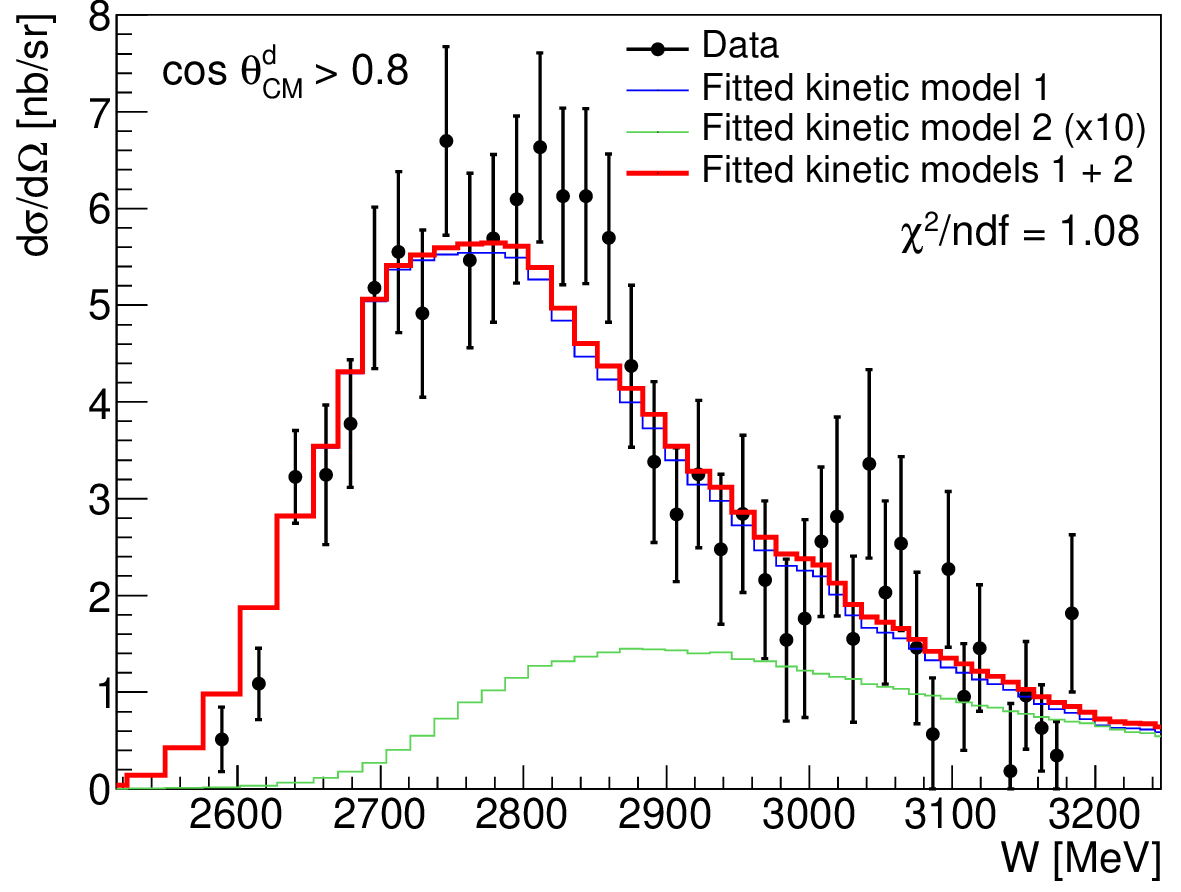}
	\caption{KM1 and KM2 shown in Fig.~\ref{fig:coherentdiagrams} fitted to  the $\gamma d \rightarrow \pi^0\eta d$ differential cross section.  The thick red  line is the sum of KM1 and KM 2 (blue and green respectively), achieving a reduced $\chi^2$ of 1.08.  KM2 is scaled by a factor of ten for visibility.}
	\label{fig:coherentfit}
\end{figure}
    
 Figure~\ref{fig:invmass} shows the $\pi^0 d$, $\eta d$ and $\pi^0\eta$ invariant mass distributions for three intervals of $W$ which span the measured differential cross section range.  Five different model distributions are superimposed; a phase space distribution (shaded green region), KM1 (thick red line) and two models of a bound two-baryon intermediate state   with sequential meson emission,  shown in Fig.~\ref{fig:coherentdiagrams}(c) and (d).  Both of these models are similar to  KM1 and KM2 in that on-shell kinematics are used and the amplitudes of the distributions are normalised to the data.  The blue line assumes a bound $\Delta(1232) N$ system following $\eta$ emission, with a subsequent $\pi^0$ decay of the $\Delta(1232)$ and the magenta line assumes a bound $N(1535)N$ system following $\pi^0$ emission with a subsequent $\eta$ decay of the $N(1535)$.    The thick purple line (where available and shown at an arbitrary scale) is the phenomenological model of Ishikawa et al.~\cite{ishikawa22} which was fitted to  data in Ref.~\cite{ishikawa22} integrated over all \dangle{}. This model assumed an amplitude of two terms:  The first was a quasi-two-body state with $J^P = 2^+$ and often described as a $\Delta N$ dibaryon.  The second was a virtual $\eta d$ state residing close to the $\eta d$ threshold.

  It is clear that phase space alone does not describe the data, with the exception of close to threshold, where limited phase space makes it hard to discern between distributions. 
 

 \begin{figure*}[h]
	\includegraphics[trim={4cm 10cm 0cm 0cm},clip,width=\textwidth]{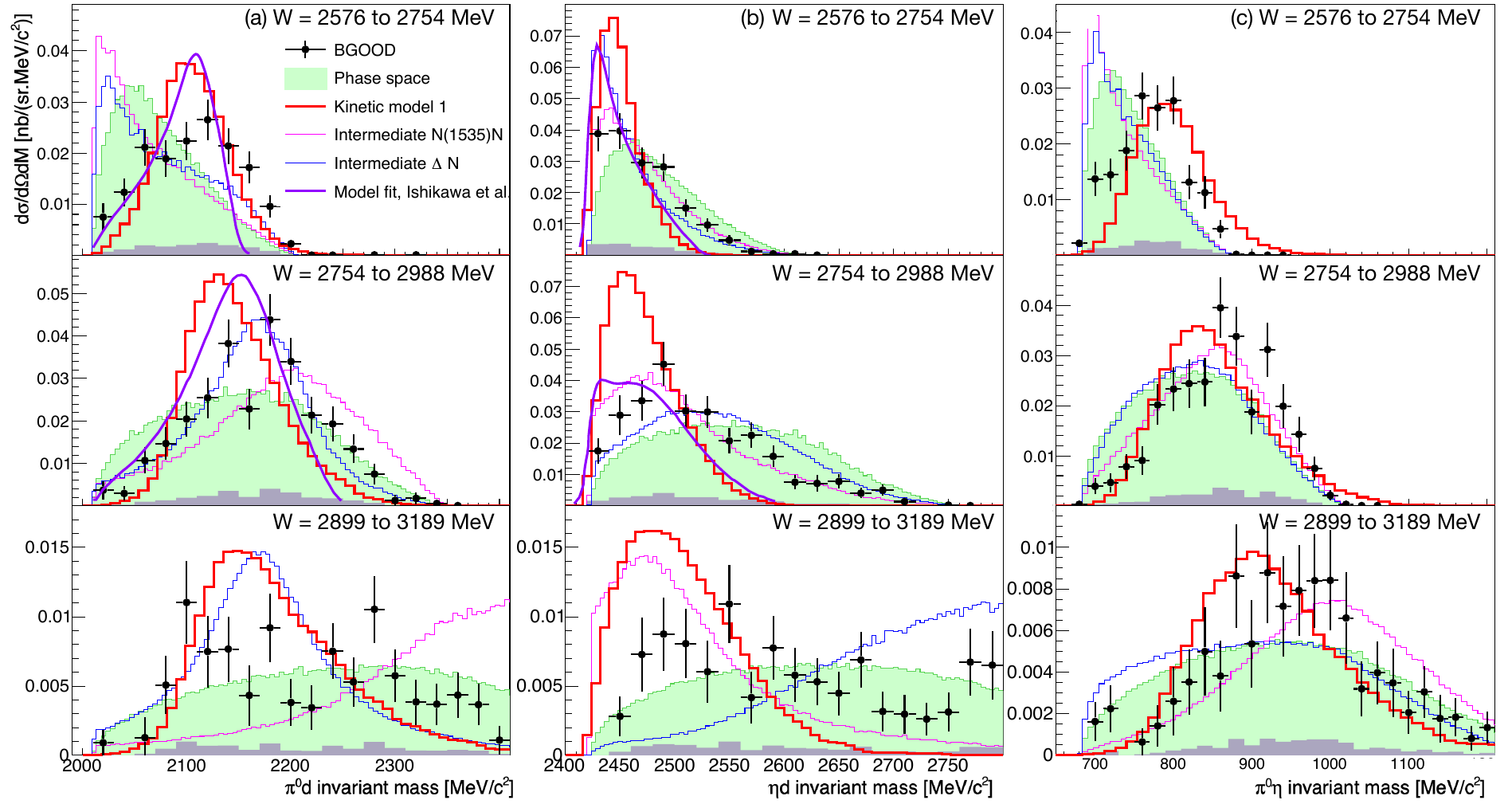}
	\caption{The differential cross section versus the invariant mass for the (a) $\pi^0d$, (b) $\eta d$ and (c) $\pi^0 \eta$ systems for \dangle{} $> 0.8$.  Each row corresponds to a different $W$ range, labelled inset (note the difference in scale of the y-axes).  The data are shown as filled black circles, where the vertical error bar is the statistical error and the horizontal bar the mass interval.  The systematic uncertainties are indicated by the grey bars on the absisca.  The different model distributions are included, with the integrals equal to the experimental data.  The shaded green region is phase space, the thick red line is Kinetic Model 1  (described in the text and shown in Fig.~\ref{fig:coherentdiagrams}(a)), the blue line is assuming an intermediate $\Delta N$ bound system and the magenta line is assuming an intermediate $N(1535)N$ bound system.  The thick purple line  (where available) is the phenomenological model of Ishikawa et al.~\cite{ishikawa22} and fitted to the data integrated over all \dangle{} in the same reference.  The distribution is plotted here at an arbitrary scale.  The two intervals in $W$ are not the same as labelled, but are instead 2661 to 2703\,MeV (top panels) and 2703 to 2799\,MeV (middle panels).}
	\label{fig:invmass}
\end{figure*}

  \subsection{The $\pi^0 d$ invariant mass distributions}
  
  The KM1 and the model of Ishikawa et al.~give reasonable descriptions of the $\pi^0 d$ invariant mass at lower $W$.   In the highest $W$ interval and with limited statistical precision, the data has peaks around 2100\,MeV/c$^2$ which would be consistent with the proposed isovector dibaryon candidate observed in the $\pi^0\pi^0 d$ channel with the FOREST at ELPH detector with a mass of 2.15\,GeV/c$^2$ and a width of 90\,MeV/c$^2$~\cite{ishikawa19}.\footnote{The BGOOD publication for the forward $\gamma d \rightarrow \pi^0\pi^0 d$ differential cross section found structure consistent with this state in the $\pi^0 d$ invariant mass~\cite{coherentpaper}, however the measured width was no broader than the experimental resolution of approximately 20\,MeV/c$^2$.}  
  
  \subsection{The $\eta d$ invariant mass distributions} 

The data for the $\eta d$ invariant mass appears to have a low mass enhancement, preferring the distributions of KM1 and the bound $N(1535)N$ model over the phase space distribution.  It is also in reasonable agreement with the phenomenological model of Ishikawa et al.~\cite{ishikawa22}, which included a large $\eta d$ scattering length and a virtual  $\eta d$ state. The agreement between models would be expected due to the subsequent $\eta N$ decay of the $N(1535)$ with only a small phase space available.  The relative momentum between the $\eta$ and the deuteron appears too large for it to be likely that a bound $\eta d$ system is the origin of this low mass enhancement.  Even close to threshold where it is lowest, the relative momentum distribution in the lab frame is approximately Gaussian, with a mean of 526\,MeV/c and a $\sigma$ of 56\,MeV/c.

  \subsection{The $\pi^0 \eta$ invariant mass distributions} 

 KM1 and  the bound $N(1535)N$ model appear to give a qualitatively reasonable description of the $\pi^0\eta d$ invariant mass.  It was observed that data taken at the FOREST detector at ELPH could be described by a plane wave description of the invariant mass distribution, which would be expected due to the small $\eta \pi$ interaction below the $a_0(980)$ threshold at $W = 2855$\,MeV~\cite{ishikawa22}.  This BGOOD data set exceeds the $a_0(980)$ threshold in the highest $W$ interval, however there does not appear to be any structure which could be attributed to the $a_0(980)$, and  KM1 still gives a reasonable description.
 
\section{Conclusions}

The  coherent reaction, $\gamma d \rightarrow  \pi^0\eta d$ was studied with high statistical precision at forward deuteron angles for the first time.  This was achievable due to the excellent particle identification of the BGOOD Forward Spectrometer to separate deuterons and protons from quasi-free reactions.

The differential cross section is orders of magnitude higher than expected from phenomenological models which assume the impulse approximation for the elementary process, $\gamma N\rightarrow \Delta(1700)\rightarrow N(1535)\pi^0\rightarrow \pi^0\eta N$ and include the deuteron momentum form factor and meson re-scattering terms.  The large momentum transferred to the deuteron in this kinematic regime is significantly higher than the internal Fermi momentum, which would prevent the deuteron to remain in tact and for a conventional coherent reaction to occur.  The distribution of the differential cross section is excellently described by the ``toy model" depicted in Fig.~\ref{fig:coherentdiagrams}(a), involving quasi-free production of a $\Delta(1232)$, $\pi$ rescattering to an $N(1535)$ and a subsequent coalescence of the nucleons to a deuteron.  The model also gives a reasonable description of the invariant mass distributions and the similar decay branching ratios of the $N(1535)$ to $\eta N$ and $\pi N$ would  explain the similar strength of the cross section to $\gamma d \rightarrow \pi^0\pi^0 d$.  A  quantitative calculation beyond the scope of this paper would be desired to confirm this.  If this reaction is indeed dominated by such pion-rescattering and nucleon coalescence mechanisms, this may disfavour the role of candidate dibaryons contributing to $\gamma d \rightarrow \pi^0\pi^0 d$ and other reactions with baryon number 2 intermediate systems.

\section*{Acknowledgements}

We thank the staff and shift-students of the ELSA accelerator for providing an excellent beam.  We would also like to thank T.~Ishikawa for providing data from Ref.~\cite{ishikawa22}.   

This work is supported by the Deutsche Forschungsgemeinschaft Project Numbers 388979758 and 405882627 and the Third Scientific Committee of the INFN.   This publication is part of a project that has received funding from the European Union’s Horizon 2020 research and innovation programme under grant agreement STRONG–2020 No.~824093. P.~L.~Cole gratefully acknowledges the support from both the U.S. National Science Foundation (NSF PHY-1307340, NSF-PHY-1615146, and NFS-PHY-2012826) and the Fulbright U.S. Scholar Program (2014/2015).

\bibliographystyle{unsrt}
\bibliography{/Users/tom/Documents/HabilitationReferences.bib}

%
%
%

\end{document}